\pdfoutput=1
\documentclass[epj-spec]{svjour}
\usepackage[english]{babel}
\usepackage[ansinew]{inputenc}
\usepackage[T1]{fontenc}
\usepackage{amsmath}
\usepackage{amssymb}
\usepackage{amsfonts}
\usepackage{graphicx}
\usepackage{color}
\usepackage{latexsym}
\usepackage{bm}

\renewcommand \thesection{\Roman{section})}

\begin{document}

\title{The 3-Band Hubbard-Model versus the 1-Band Model for the high-$T_c$ Cuprates: Pairing Dynamics, Superconductivity
and the Ground-State Phase Diagram}

\author{W. Hanke  \inst{1} \and M.L. Kiesel  \inst{1} \and M. Aichhorn \inst{2} \and S. Brehm \inst{1} \and E. Arrigoni  \inst{3}}
\institute{Institute for Theoretical Physics, University of W\"urzburg, Am
Hubland, 97074~W\"urzburg, Germany \and Centre de Physique Th\'{e}orique, \'{E}cole Polytechnique, CNRS,
91128 Palaiseau Cedex, France \and Institute of Theoretical and Computational Physics, Graz University
of Technology, Petersgasse 16, 8010 Graz, Austria}

\abstract{One central challenge in high-$T_c$ superconductivity (SC)
is to derive a detailed understanding for the specific role of the
$Cu$-$d_{x^2-y^2}$ and $O$-$p_{x,y}$ orbital degrees of freedom. In most
theoretical studies an effective one-band Hubbard (1BH) or t-J model
has been used. Here, the physics is that of doping into a
Mott-insulator, whereas the actual high-$T_c$ cuprates are doped
charge-transfer insulators. To shed light on the related question,
where the material-dependent physics enters, we compare the
competing magnetic and superconducting phases in the ground state,
the single- and two-particle excitations and, in particular, the
pairing interaction and its dynamics in the three-band Hubbard (3BH)
and 1BH-models. Using a cluster embedding scheme, i.e. the
variational cluster approach (VCA), we find which frequencies are
relevant for pairing in the two models as a function of interaction
strength and doping: in the 3BH-models the interaction in the low-
to optimal-doping regime is dominated by retarded pairing due to
low-energy spin fluctuations with surprisingly little influence of
inter-band (p-d charge) fluctuations. On the other hand, in the
1BH-model, in addition a part comes from ''high-energy'' excited
states (Hubbard band), which may be identified with a non-retarded
contribution. We find these differences between a charge-transfer
and a Mott insulator to be renormalized away for the ground-state
phase diagram of the 3BH- and 1BH-models, which are in close overall
agreement, i.e. are ''universal''. On the other hand, we expect the
differences - and thus, the material dependence to show up in the
''non-universal'' finite-T phase diagram ($T_c$-values). }

\maketitle

\section {Introduction}
Many aspects of the physics of high-temperature
superconductors (HTSC) remain mysterious, despite impressive
progress both on the experimental and theoretical front. One key
issue, is why are the HTSC materials composed out of $CuO_2$ planes
and what is the specific role of $Cu$-$d$ and $O$-$p$ orbital degrees of
freedom? The cuprate materials in the undoped, i.e. ``half-filled''
situation, are ``charge-transfer'' insulators \cite{ref:1,ref:2}. This
fact induces an experimentally observed asymmetry between hole ($h$)-
and electron ($e$)-doping: while doped holes go onto $O$ orbitals and
may be bound to $Cu$ spins to form ``Zhang-Rice'' singlets, doped
electrons reside mainly on the $Cu$ orbitals
\cite{ref:3,ref:4,ref:5,ref:6,ref:7,ref:8,ref:9,ref:10}. This is
believed to be intimately related to the more extended stability of
antiferromagnetic (AF) behavior as a function of electron doping
compared to that of hole doping. While introducing electrons on the
$Cu$ sites merely produces a dilution of spins, holes on $O$ sites
create a ferromagnetic coupling between neighboring $Cu$ orbitals,
which is significantly more effective in destroying the AF order.
Beyond magnetism, the asymmetry is also exposed in the
superconducting (SC) behavior with the $h$-doped materials
exhibiting SC usually over wide doping regimes and with high $T_c$'s
up to $150$K, whereas in the $e$-doped system $T_c$ is low and
confined to a very narrow doping regime.

In much of the high-$T_c$ theoretical studies the starting point has
not been the above three-band Hubbard (3BH) but instead the one-band
Hubbard (1BH) and $t$-$J$ models \cite{ref:11,ref:3}. Here, the oxygen
degrees of freedom are eliminated approximatively via the Zhang-Rice
construction \cite{ref:3}. The mapping has as a crucial consequence
that the undoped model systems are Mott insulators and no longer
governed by the charge-transfer energy $\Delta_{pd}$ between the
$Cu$ and $O$ orbitals. Analytical and, in particular, numerical
calculations based on these two-dimensional (2D) single-band models
have demonstrated Fermi surfaces, single-particle spectral weights,
AF spin correlations and $d_{x^2-y^2}$ pairing correlations in
qualitative agreement with experimental measurements
\cite{ref:12,ref:13,ref:14,ref:15,ref:16,ref:17,ref:18,ref:19,sc.06}. This
fact has significantly contributed to the wide-spread belief that
the physics of HTSC is that of ``doping into a Mott insulator''
\cite{ref:20}. However, how can this picture be reconciled with the
charge-transfer insulator picture embedded in the 3BH-model? Can it
be that the charge-transfer energy $\Delta_{pd}$ in the 3BH-model,
the size of which is already decisive for the accuracy of the 1-band
reduction \cite{ref:3}, plays the role of an effective on-site
Hubbard $U$ in the 1-band models?

Quantifying these ideas requires solving the strongly correlated
electron problem for the 3BH- and 1BH-models at very low energy
(and/or temperatures). Early Quantum-Monte-Carlo (QMC) calculations
for the 3BH-model showed, that characteristic features such as the
doping dependence of the electronic single-particle excitations and
their interplay with magnetic excitations are in accord with
experiment \cite{ref:4,ref:5}. However, the very low $T$- or
ground-state properties including the SC state, could not reliably
be resolved, due to the well-known ``minus-sign'' problem
\cite{ref:21}. Embedded cluster techniques provide a controlled way
to approach the infinite-size (and, thereby, low-energy) limit.
Recently, the variational cluster approach (VCA) which was proposed
by M. Potthoff and our group \cite{ref:22,ref:23} has been shown for
the 1BH model to correctly reproduce salient features of the
ground-state ($T=0$) phase diagram of the high-$T_c$ cuprates
\cite{ref:14,ref:16,ref:17,ar.ai.09}. In particular, the AF and d-wave SC
ground-states were found in doping ranges qualitatively in accord
with experimental data for both $e$- and $h$-doping such as the
different stability of the AF phase. It can be accounted for by a
``simple'' 1BH model, in which the $e$-$h$ symmetry is broken by a
longer-ranged (next-nearest) hopping term.

Regarding the possibility of the reduction to a one-band model,
there are questions concerning the direct applicability of the
Zhang-Rice (ZR) construction. As discussed in the literature before
\cite{ref:9}, the natural tendency of a finite oxygen band width is
to delocalize and to destabilize the ZR singlets. Secondly, the
pragmatic finding that a $t$-$t'$-$U$ 1BH-model with a significant
value of $t'$, captures basic physics of the cuprates and in
particular their $e$-$h$ asymmetry, cannot be accounted for in a
strict ZR picture \cite{ref:24}. In this picture, next-nearest
neighbor hoppings are very small compared to nearest-neighbor terms
(if again oxygen-oxygen hopping $t_{pp}$ is taken into account).
What we will show in our calculations is that the 3BH-model and a
single-band $t$-$t'$-$U$ Hubbard model with a significant value for
$t'$ (which may be taken as an empirical parameter adjusted to fit
the Fermi-surface topology \cite{ref:25}) exhibit a similar
low-energy (here specifically, $T=0$) physics concerning the
qualitative behavior of the ground-state phase diagram \cite{25a}
and of the single-particle excitations including the asymmetry as a
function of $h$- and $e$- doping.

Then, where do the cuprate material-specific properties appear? To
shed light on this question, we start out in section III with
contrasting the low-$T_c$ pairing mechanism with a discussion of
what we know about the high-$T_c$ mechanism. In particular, we take
up an issue raised recently by P.W. Anderson \cite{ref:28}, whether
the pairing interaction in the cuprate HTSC should be considered as
arising from a ''pairing glue''. Anderson argued that there is no
room left for a pairing glue: while the low-$T_c$ SC pairing is due
to the well-known dynamic screening mechanism which contains a
(phononic) ''pairing glue'', in the high-$T_c$ cuprates and other
unconventional SC another mechanism is at work. It may be termed
''anisotropic real- or momentum-space'' pairing. Here, the electrons
pair in an anisotropic wave function (such as d-wave), which
vanishes at the repulsive core of the Coulomb interaction. While
this anisotropy is certainly realized for the cuprate HTSC and
embedded in the momentum dependence of the d-wave gap function,
there is, additionally, a frequency dependence of the SC gap, which
tells us about the dynamics of the pairing interaction
\cite{ref:26}. We study in section IV the dynamics of the SC gap
function and the building up of a ``pairing glue'' which contributes
to the formation of Cooper pairs in the 3BH-model and contrast it
with the corresponding ``pairing glue'' for the 1BH-model. The
``pairing glue'' question is a natural one to ask if one has
two-particle excitations, which are responsible for the dynamics of
the pairing interaction, whose energy scale is small and of the
order of the SC gap energy. For the 1-band models this ``pairing
glue'' issue has recently been investigated
\cite{ref:26,ref:31,31b}. Spin fluctuations create a retarded
interaction, which is dominating the  low-energy pairing. However,
also a pairing contribution due to interband (lower and upper
Hubbard band) transitions was identified \cite{ref:26} in the
1BH-model case. In our calculation for the 1BH-model, this
''non-retarded'', i.e. spin-fluctuation dominated part is comparable
in magnitude to a second part: For large $U$'s this creates an
``instantaneous'' pairing interaction eventually going over into the
instantaneous part previously identified  in the $t$-$J$ model
\cite{ref:26}.

There are even qualitative differences in the dynamics of the
pairing interaction in the 3BH-model: here, the dynamics of the
pairing interaction is dominated by the retarded contribution due to
spin fluctuations. Higher-energy p-d charge fluctuations give rise
to a surprisingly small further (retarded) contribution. As
expected, this charge-fluctuation contribution gains slightly more
weight, when going to higher hole dopings. In short, one may term
the dynamics of the pairing interaction in the 3BH-model and, thus,
in a doped charge-transfer insulator as retarded and being due to a
low-energy pairing glue, whereas in the 1BH-model and, thus, a
doped Mott insulator both retarded and non-retarded parts
contribute. The differences in the dynamics of the pairing
interaction, which are due to the material dependent
$CuO_2$-physics, appear at relatively large energies (of order
$\Delta_{pd}$) which are significantly larger than the SC gap. In
the ground-state phase diagram these ``high-energy'' differences are
renormalized away (see, Fig. \ref{Figure 5}, in section IV). We do,
however, expect that they play a role in the finite-T phase diagram
($T_c$-values).

Also the low-energy single-particle excitations of the 3BH- and
1BH-model are qualitatively similar, when comparing the 3BH-model
with the (as discussed above) empirical t-t'-U model. A detailed
discussion of this was presented in Ref. \cite{ar.ai.09}. This
concerns, in particular, where holes or electrons, respectively, go
when one dopes away from half-filling. The corresponding nodal and
anti-nodal doping regimes are directly related to the observed
asymmetry in the robustness of the AF order. Closer inspection,
however, also reveals differences for higher energies of $\mathcal
O(t_{pd})$. An example is the recently much discussed ``waterfall''
structure or high-energy kink appearing as an abrupt change in the
band dispersion, which falls vertically at binding energies below
$\approx0.4eV$ \cite{ref:27}. This ``waterfall'' behavior is found
to be rather pronounced in the 3BH model, but not in the 1BH-model.

Finally, in section V, we describe a theory we have recently
developed of two-particle (2-p) excitations and, in particular, the
magnetic susceptibility in high-$T_c$ cuprate superconductors
\cite{br.ar.10}. As just discussed, 2-p excitations such as spin and
charge excitations. play a key role in the pairing theory. There,
they are indirectly embedded in the anomalous part of the selfenergy
and the dynamics of the gap function. However, for an unambiguous
identification, for example of a ''pairing glue'', one obviously
needs a direct scheme for calculating these 2-p excitations in the
experimentally relevant strong-correlation regime. Therefore, we
have recently extended the Variational Cluster Approach to these
excitations \cite{br.ar.10}. We found for the 1BH-model, that the
corresponding magnetic susceptibility $\chi_s$ reproduces salient
features of the experiment such as the celebrated neutron-scattering
resonance in the hole-doped case. Previous descriptions of the
magnetic resonance have been obtained by weak-coupling
\cite{is.er.07} and/or semiphenomenological approaches \cite{ref:43}
and, quite independently, by the SO(5)-symmetry argument
\cite{ref:44}. The infinite-lattice limit is crucial to obtain the
magnetic resonance which may be
 considered as a ”fingerprint” of the AF order in the SC state. Only then are
we able to differentiate between the competing AF and SC orders in
the phase diagram. Therefore, this limit, which is obtained with the
VCA, has also to be embedded in a controlled description of the
corresponding susceptibilities.

In section V, we also study the emergence of a similar magnetic
resonance in the SC state of the electron-doped cuprate
superconductors. We find again for the 1BH-model case, that the
experimentally observed resonance peak is consistent with an
overdamped magnetic (spin S=1) exciton located near the
particle-hole continuum. Because of computational reasons (reference
cluster size) we can only demonstrate the magnetic resonance for the
1BH-model.


\section {The Models}
We consider the following version of the three-band Hubbard model:

\begin{equation}
\begin{split}
H=
& \sum_{\sigma,<ij>}t_{ij} \left( d^\dagger_{i\sigma}p_{j\sigma} + H.c.\right)
+\sum_{\sigma,<jj'>}\bar t_{jj'}\left( p^\dagger_{j\sigma}p_{j'\sigma} + H.c.\right)
+\left(\epsilon_d -\mu\right)\sum_{\sigma,i}n^d_{i\sigma}\\
&+\left(\epsilon_d -\mu\right)\sum_{\sigma,j}n^p_{j\sigma}
+U_{dd}\sum_i n^d_{i\uparrow} n^d_{i\downarrow}
+U_{pp}\sum_i n^p_{i\uparrow} n^p_{i\downarrow}
+U_{pd}\sum_{\sigma\sigma',<ij>} n^d_{i\sigma} n^p_{j\sigma'}
\end{split}
\end{equation}

where $d_{j\sigma}^\dagger$ creates a hole with spin $\sigma$ in the
$Cu$-$d_{x^2-y^2}$ orbital at site $i$ with occupation number
$n_{i\sigma}^d$. Correspondingly, $p_{j\sigma}^\dagger$ creates an
$O$-$p_{x,y}$ hole at site $j$ with occupation number
$n_{j\sigma}^p$. $t_{ij}$ stands for the p-d hybridization
$t_{Cu-O}$ and $\overline{t}_{jj'}$ for the direct $O$-$O$ hopping,
where the orbital phase factors are included \cite{ref:4,ref:5}. The
local orbital levels are given by $\epsilon_p$ and $\epsilon_d$ and
the charge-transfer energy is $\Delta=\epsilon_p-\epsilon_d$.
$U_{dd}$ and $U_{pp}$ are the Hubbard couplings on the $Cu$ and $O$
sites, respectively. Finally, $U_{pd}$ is a repulsive interaction
for holes on $Cu$ and $O$. We take typical parameters consistent
with earlier extensive Quantum-Monte-Carlo (QMC) \cite{ref:4,ref:5}
as well as cluster \cite{ref:25} calculations. In units of the
$Cu$-$O$ hopping $t_{pd}=1$, $\Delta=3$, $U_{dd}=8$, $U_{pd}=0.5$,
$U_{pp}=3.5$ and $t_{pp}=0.5$.

The 1BH-model is defined as usual, i.e.

\begin{equation}
H=
\sum_{\sigma,<ij>}t_{ij}c^\dagger_{i\sigma}c_{j\sigma}+U\sum_i n_{i\uparrow}n_{i\downarrow}.
\end{equation}

Here $c_{j\sigma}$ and $c_{i\sigma}^\dagger$ are annihilation and
creation operators and $t_{ij}$ denote the nearest ($t_{nn}=1$,
energy unit) and next-nearest neighbor ($t_{nnn}=-0.3$) hopping
matrix elements and $U$ $(U=8)$ the on-site Hubbard repulsion.

An important remark for the comparison of the two models is that while for the 3BH-model the unit is $t_{pd}=1$ and $t_{nn}=t=1$ for the 1BH-model, when going back to eV's one has to set $t_{pd}\cong1.5[eV]$ and $t\cong0.4[eV]$.

\section {Low-$T_c$ versus high-$T_c$ superconductivity and the question of the pairing
glue.}

In a recent article \cite{ref:28}, P. W. Anderson (PWA) has argued
that pairing in conventional ``low-$T_c$'' superconductors (SC) has
a rather different microscopic origin from that in high-$T_c$
cuprates and many other unconventional SC. In either case, the
paired electrons have to avoid the strongly repulsive bare Coulomb
interaction.

In a \emph{low-$T_c$ SC} this repulsion can be eliminated in favor
of electron-pair binding via \emph{``dynamic screening''}, i. e.
\begin{equation}
V(\vec{q}, \omega)=\frac{e^2}{q^2\varepsilon(\vec{q}, \omega)},
\end{equation}
where $V(\vec{q}, \omega)$ is the Fourier transform of the el-el
interaction in both space and time and $\varepsilon(\vec{q},
\omega)$ is the dynamic screening due to both ions and electrons,
i.e. $\varepsilon({q}, \omega)=\varepsilon_{ion}(\vec{q},
\omega)+\varepsilon_{el}(\vec{q}, \omega)$.

PWA then gives an elegant discussion why, in his opinion, in the
\emph{high-$T_c$ SC} another pairing mechanism is at work, which may
be termed \emph{``anisotropic momentum (or real-space) mechanism''}:
here, the strongly repulsive (short-range) part of the Coulomb
interaction is avoided by a mechanism suggested by Pitaevskii
\cite{ref:29} and Brueckner et. al. \cite{ref:30} of choosing the
pair state orthogonal to the repulsive core of the Coulomb
interaction, i.e. putting the electron pairs in an anisotropic wave
function (such as d-wave), which vanishes at the core of the Coulomb
interaction.

While we follow the arguments of PWA for the low-$T_c$ SC, we would
like to present here evidence, that \emph{in the high-$T_c$
materials both the anisotropic ``momentum space'' mechanism and the
dynamics} are \emph{at work}. In what follows this will be shown for
both of the above Hubbard-type of  models (see chapter IV). This
gives us, in particular, a possibility to shed light on the question
to what extent both models lead to a similar pairing mechanism. For
the 1-band Hubbard model, the reader is also referred to recent
studies by Maier, Poilblanc and Scalapino \cite{ref:26} as well as
by Kyung, S\`{e}n\`{e}chal and Tremblay \cite{ref:31}, where also
the pairing dynamics has been studied. The above questions are
clearly of relevance for the main issue raised by PWA, namely ``is
there glue in cuprate superconductors?'', which is basically a
question about the dynamics of the pairing interaction. As argued by
Maier et al., if the dynamics of the pairing interaction arises from
virtual states, whose energies correspond to the ``high-energy''
Mott gap, and give rise to the exchange coupling $J$, the
\emph{interaction is instantaneous on the relative time scales of
interest}. However, if the energies correspond to typical
``low-energy'' spin-fluctuation (or phonon) excitations, then the
\emph{interaction is retarded}. In this case it makes sense to use
the terminology ``spin-fluctuation glue'', which mediates the d-wave
pairing. Here, we present results from numerical (variational
cluster) studies, which provide insight into this question.

For this presentation it is useful to step by step \emph{contrast}
the construction of the \emph{effective el-el interaction for a
weak-coupling low-$T_c$ SC with the effective interaction of a
high-$T_c$ (i.e. Hubbard type) SC}. In the ``dynamic screening''
mechanism of eq. (3) one can safely replace the electronic screening
by a static one (i.e. $\varepsilon_{el}(\vec{q},\omega=0)$). This is
due to the fact that typical el-energies are of ``high energy''
$(\sim \mathcal O(\varepsilon_{Fermi})$ compared to $k_BT_c$, i.e.
the SC energy scale: the plasma of other electrons then damps away
the long-range ($1/r$)-behavior and leaves a Thomas-Fermi screened
core $e^2e^{-xr}/r$ ($x$: Thomas-Fermi constant). This gives rise to
an essentially instantaneous interaction, which is still ``very
repulsive'', and which - when averaged over the Fermi surface - is
termed $\mu$ (see below). On the other hand, for the phonon case,
the screening acts anti-adiabatically, i.e.
$\varepsilon_{ion}(\vec{q}, \omega)$ is dynamic, since typical
phonon frequencies are of the order of $k_BT_c$. In other words, the
final Fourier-transformed effective interaction is
$V_{\text{eff}}=\frac{e^2}{[(q^2+x^2)\varepsilon(q,\omega)]}$, and
the effective electronic interaction is screened
(anti-adiabatically) by the phonon polarizations.

\begin{figure}
\begin{center}
   \includegraphics[width=0.8 \columnwidth]{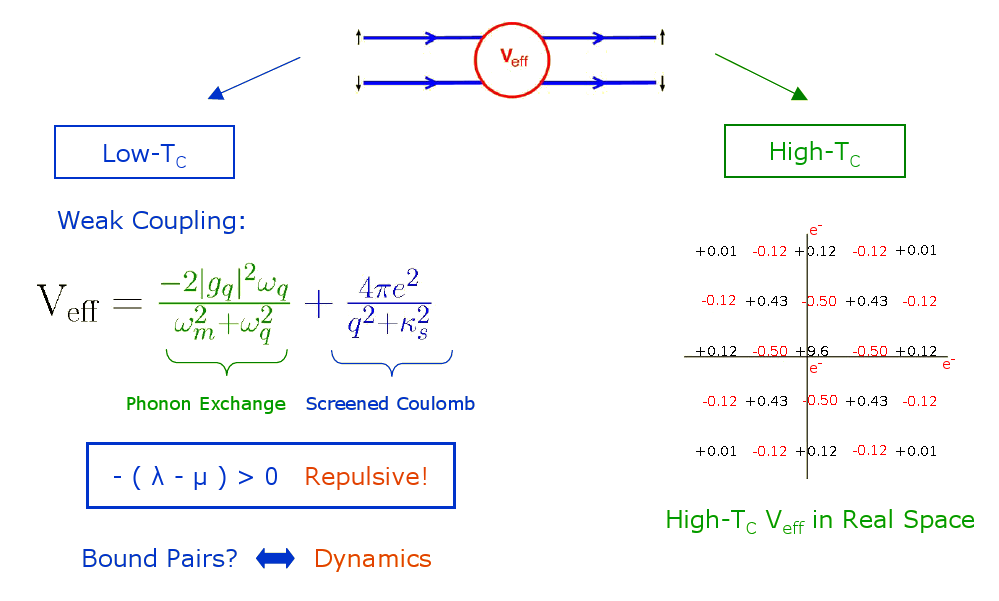}\\
  \caption{The effective pairing interaction in the weak-coupling low-$T_c$ situation compared with high-$T_c$ QMC simulations for
the 2D 1-band Hubbard model by the Scalapino group \cite{ref:35}. }
  \label{Figure 1}
\end{center}
\end{figure}

\begin{figure}
\begin{center}
   \includegraphics[width=0.9 \columnwidth]{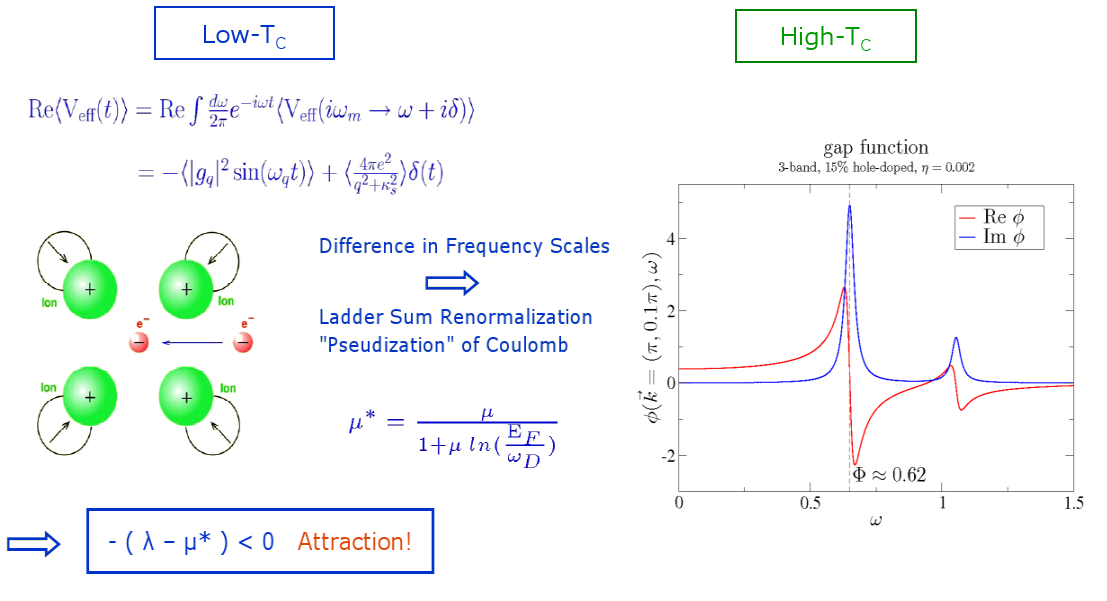}\\
  \caption{left-hand part: Dynamics of the pairing interaction for the low-$T_c$ case: the emergence of the difference in frequency scales, i.e. slow ionic-lattice polarization and fast Coulomb repulsion. Right-hand part: the SC gap function and its dynamics (both Re- and Im- parts) for the 3BH-model. Note that the Re-part displays a certain similarity to the gap function shown for the low-$T_c$ case in the left-hand part of Fig. \ref{Figure 3}. This is true for the spin-fluctuation contribution appearing around the characteristic energy $\omega_0 + \Delta_0\cong\Phi$ (see also Fig. \ref{Figure 6}f)). At higher energies of order $\omega\cong1$ and above additional structure appears mainly due to $p-d$ charge fluctuations (see again Fig. \ref{Figure 6}f)).}
  \label{Figure 2}
\end{center}
\end{figure}

Fig.~\ref{Figure 1} summarizes this weak-coupling low-$T_c$
situation and rewrites the effective pairing interaction as the
usual sum of the dynamic phonon exchange and the Thomas-Fermi
screened Coulomb part. Here $\omega_q$ denotes phonon (ph)
frequencies, $g_{\vec{q}}$ the el-ph coupling and $\omega_m$ are
Matsubara frequencies. When the two terms in the sum are averaged
over the Fermi surface (the brackets in Fig.~\ref{Figure 2} for
$V_{\text{eff}}$ denote an average of the momentum transfer q over
$\varepsilon_{F}$), then one finds that the corresponding averaged
$V_{\text{eff}}$ is larger than zero, i.e. -$(\lambda-\mu)>0$ and,
thus, is still repulsive. The net interaction is thus repulsive even
in the phonon case, a fact which is required to guarantee the
stability of the solid.

So, how can one ever arrive at bound pairs, if the interaction is
never attractive? As is well-known since Cooper's seminal paper
\cite{ref:32} preceding BCS-theory, this is due to the difference in
frequency scales or, equivalently, in time scales of the two parts
of the interaction. This is pictured in Fig.~\ref{Figure 2}: the ``first''
electron (anti-adiabatically) polarizes the ionic lattice and sets
up a net negative charge polarization in its vicinity. This first
process is ``slow'' and happens on the frequency scale $\omega_q$ of
the ionic vibrations. A ``second'' electron feels this polarization,
but can only profit from it and build up an effective attraction
when the first electron ``instantaneously'' moves so as to avoid
each other. Thus, the ``high-energy'' part of the Coulomb
interaction acts only over a short time, given here by a
$\delta$-function, while the attractive el-ph interaction is
\emph{retarded} by the much slower lattice response. In other words,
if the electrons forming the pair correlate themselves in time to
avoid the short-time Coulomb repulsion, they can take advantage of
the attractive el-ph mediated interaction and form a Cooper pair. As
is well-known, this kind of renormalization, i. e. integrating out
the ``high-energy''  degrees of freedom in the relative pair wave
function (so that one arrives at one and the same low-energy cutoff
of order of the Debye frequency $\omega_D$ in Fig.~\ref{Figure 2}),
can be done a variety of ways: by a ladder sum renormalization or a
``pseudozation'' of the interaction , replacing $\mu$ by $\mu^*$
\cite{ref:33}. Now, we can have $-(\lambda-\mu^*)<0$, i.e. a
situation where $V_{\text{eff}}$ can be attractive, leading in Cooper's
sense to the pairing instability at very low energies.

\begin{figure}
\begin{center}
   \includegraphics[width=0.8 \columnwidth]{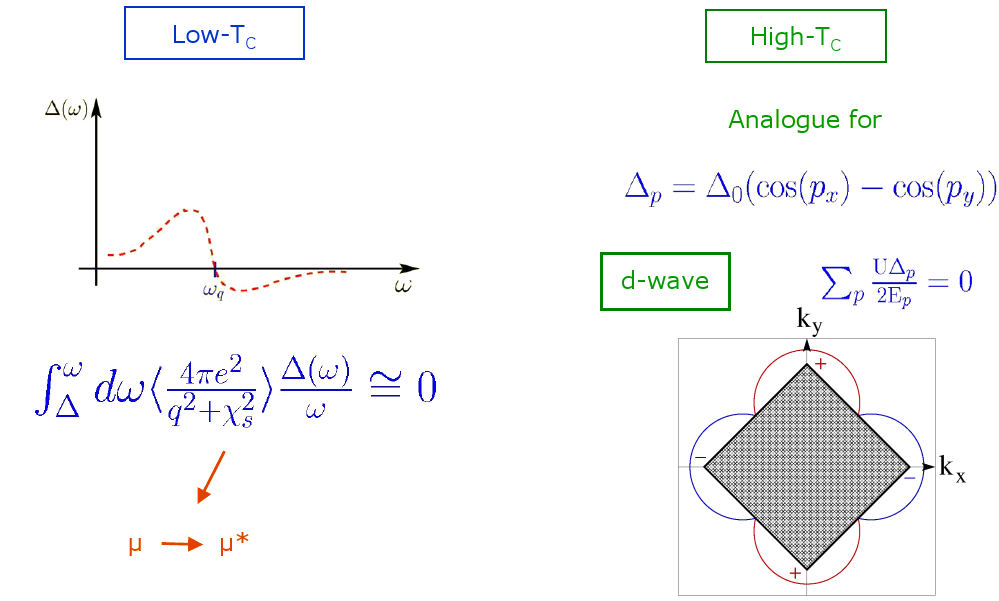}\\
  \caption{Avoiding the strongly repulsive part of the Coulomb repulsion: Analogy of the gap function $\Delta$ in the low-$T_c$ $(\omega-space)$ and high-$T_c$ (momentum space) situation. The figure for the momentum-space gap function in the high-$T_c$- case is again taken from \cite{ref:35}. Here, the shaded region is the normal-state Fermi surface and the solid curve surrounding the Fermi surface gives the size of the $d_{x^2-y^2}$-gap $\Delta_k$ at the momentum k on this surface.}
  \label{Figure 3}
\end{center}
\end{figure}

We are interested in the dynamics of the pairing interaction in the
\emph{high-$T_c$ cuprates} and, in particular, how there the two
electrons about to form a pair can avoid each other - and thus
weaken their repulsion - by modifying the high-energy parts of their
relative wave function. For this, it is instructive to look first at
the frequency dependence of the gap function of a low-$T_c$ SC found
by solving the Eliashberg equations \cite{ref:34}, following a
discussion by D. J. Scalapino \cite{ref:35} (see the low-$T_c$ case
in Fig.~\ref{Figure 3}): The real part of the corresponding gap
function $\Delta(\omega)$ first increases as the typical phonon
energy $\omega_{\vec{q}}$ is approached. At this characteristic
``glue energy'', it changes sign and remains negative out to very
large values of $\omega$. This latter observation corresponds to the
instantaneous part. It just reflects the fact that the two electrons
making up the pair avoid ``short-time close-range encounters''. This
can be summarized (see Fig.~\ref{Figure 3}) in a kind of
orthogonality relation, with the pair wave function $\Delta(\omega)$
being orthogonal to the ``core'' (i.e. the Thomas-Fermi screened
short-range part) of the Coulomb interaction \cite{ref:35}. The
essence of this orthogonality relation is that in practice
$\langle4\pi e^2/(q^2+x^2)\rangle$ can be replaced over the
frequency integral $\Delta$ to several times the Debye frequency
$\omega_D$ by the weak pseudopotential $\mu^*$ \cite{ref:33}
\cite{ref:35}.

So then, how do the electrons in high-$T_c$ materials act so that
they are seldom or never in the same place at the same time?

Here, it is useful to first look at the effective interaction
$V_{\text{eff}}$ in real-space (or, more precisely, at the
real-space Fourier transform of the singlet vertex
$\Gamma(l,\omega_m=0)$ \cite{ref:34} versus the separation l between
the electrons in pairs). Figs.~\ref{Figure 1}~\&~\ref{Figure 3} plot
the results for $V_{\text{eff}}$ obtained from QMC simulations for
the 2D 1-band Hubbard model by the Scalapino group \cite{ref:35}. We
have obtained a similar pattern for the effective interaction in the
3BH-model employing also QMC, when again the two electrons are
placed on the Cu-lattice sites \cite{ref:36}.

Using the BCS gap equation for illustration, i.e.
\begin{equation}
\Delta_p=-\sum\limits_{p'} V_{pp'}\frac{\Delta_{p'}}{2Ep'},
\end{equation}
with $V_{pp'}$, the effective interaction or Fourier-transformed
vertex $V_{pp'}=T(p'-p)$, we immediately see that the two pairing
electrons avoid the repulsive parts of the Coulomb interaction by
arranging their pair wave-function in a $d_{x^2-y^2}$-orbit, with
the simplest nearest-neighbor pairing given in momentum space by
$\Delta_p=\Delta_0(\cos{p_x}-\cos{p_y})$. This is expressed by a
kind of Pitaevskii-Brueckner relation, i.e.

\begin{equation}
\sum\limits_{p'}\frac{U \Delta_p'}{2E_p'}=0,
\end{equation}

This latter relation indeed seems to confirm PWA's conjecture in
that the essence of the pairing mechanism in low-$T_c$ SC is
``dynamic screening'', whereas in the \emph{high-$T_c$ cuprates} it
is \emph{``anisotropic momentum space pairing''}. In Fig.
\ref{Figure 2}, we see - however - that the dynamics of the pairing
interaction as reflected in the $\omega$-dependence of the gap
function $\Delta(\omega)$, is also of relevance for the high-$T_c$
cuprates. Here again, the real part of the gap function changes sign
at a characteristic frequency $\Phi$ related to spin fluctuations
and at higher energies to $p-d$ charge fluctuations.

\section {Dynamics of the pairing interaction in the 1BH- and
3BH-models}

The question of whether one can speak of a ``pairing glue'' has
recently been addressed by Maier, Poilblanc and Scalapino for the
1BH- and $t$-$J$ models \cite{ref:26}. If the dynamics of the pairing
interaction is due to ``small-energy'' (two-particle) excitations, such
as the characteristic structures seen in the spin susceptibility,
then one might speak of the interaction as being retarded on the
relative time scales of interest and of a ``spin-fluctuation'' glue
mediating d-wave pairing. This clearly is reminiscent of the usual
phonon mediated pairing interaction in the low-$T_c$ SC
\cite{ref:26}.

It is well-known that for these latter systems the gap function
$\phi(\vec{k},\omega)$ or, more precisely, the anomalous part of the
Nambu selfenergy $\Sigma(\vec{k}; \omega)$, is only very weakly
depending on momentum $\vec{k}$, corresponding to the local
character of the pairing interaction. However, the dynamics of the
gap function is important and it enters the dispersion (Cauchy)
relations between the real and imaginary parts of
$\phi(\vec{k},\omega)=\phi_{1}(\vec{k},\omega)+i\phi_{2}(\vec{k},\omega)$,
i.e.

\begin{equation}
\phi_1(\vec{k},\omega)=
\dfrac{1}{\pi}\int \dfrac{\phi_2(\vec{k},\omega')}{\omega'-\omega} \mathrm{d}\omega'
\end{equation}

and, for $\omega=0$,

\begin{equation}
\phi_1(\vec{k},\omega=0)=
\dfrac{2}{\pi}\int \dfrac{\phi_2(\vec{k},\omega')}{\omega'} \mathrm{d}\omega'
\end{equation}

A measure for the fractional contribution to the gap function
$\phi(\vec{k},\omega=0)$ and to pairing that comes from frequencies
less than $\Omega$ can then be defined \cite{ref:26}, i.e.

\begin{equation}
I(\vec{k},\Omega)=\dfrac{\dfrac{2}{\pi}\int_0^{\Omega}\dfrac{\phi_2(\vec{k},\omega')}{\omega'} \mathrm{d}\omega'}{\phi_1(\vec{k},0)}
\end{equation}

It gives the relative contribution to the pairing of the retarded
``glue'' part and of the non-retarded, i.e. ``instantaneous'' part.
In Ref \cite{ref:26} it has been demonstrated that
$I(\vec{k},\Omega)$ is a useful diagnostic for the pairing glue
(phonon contributions) in the case of $Pb$. $I(\vec{k},\Omega)$
increases as $\Omega$ passes through $\omega_{Phonon}+\Delta_0$,
where $\Delta_0$ denotes the SC gap and $\omega_0$ the
characteristic $Pb$ phonon frequencies. Its asymptotic value
$I(\vec{k},\Omega\gg\omega_{Phonon}+\Delta_0)$ exceeds unity
reflecting the fact there exists an instantaneous Coulomb
pseudo-potential. This leads to a negative, frequency-independent
contribution $\phi_{\text{NR}}$ to $\phi(\vec{k},\omega)$ and at
high frequencies $I(\vec{k},\Omega)$ exceeds $1$ by the
``instantaneous'' contribution $-\phi_{\text{NR}}/\phi_{(0)}$. Here,
using numerical techniques, we examine the question of a ``pairing
glue'' which offers a way of distinguishing different pairing
mechanisms for the one-band and three-band Hubbard models in the
relevant strong-correlation regime for the HTSC.

The variational cluster approach (VCA) is particularly well suited
for a study of the quantity $I(\vec{k},\Omega)$. This is due to the
fact that the VCA allows for accurately calculating the real and
imaginary parts of the anomalous self-energy and, thus, of the gap
function $\phi_1(\vec{k},\omega)$. This has already been shown in a
recent letter, reproducing the experimentally found ``gap
dichotomy'' of the nodal and anti-nodal gaps in HTSC as a function
of doping \cite{ref:17}.

In the VCA for the Hubbard model, the lattice is tiled up with
(isolated) clusters of a given size. The corresponding Hamiltonian
$H'$ has the same on-site interaction $U$ as the original Hubbard
model, but it has modified single-particle hopping parameters $t'$
(and chemical potential $µ'$) with, in particular, $t'=0$ between
different clusters. The cluster provides a ``reference system'',
which spans a space of trial self-energies $\Sigma(H')$. The
self-energy that ``best'' describes the physics of the infinite
lattice is then constructed via a variational principle, searching
for the stationarity of the ground potential $\Omega(\Sigma)$ in the
subspace of the cluster self-energies $\Sigma(H')$. The trial
self-energies $\Sigma(H')$ are varied by varying the parameters
$t'$, $µ'$ and the Weiss (i.e. pairing, AF, etc.) fields and a local
chemical potential-shift term, which guarantees a consistent
treatment of the particle density \cite{ref:23}. For the SC
ground-state, a Nambu representation of the self-energy is used and
$\phi(\vec{k},\omega)$ is obtained from the $\tau_1$-component.

The results for $I(\vec{k},\Omega)$ for the 3BH-model to be
discussed below, are based on a (2$\times$2) cluster with 4 $CuO_2$
unit-cells and the self-energy (and Green´s function) of this
12-site cluster was extracted from exact diagonalization. The
geometry of our reference cluster is reproduced in Fig.~\ref{Figure
4}. As we found 2-particle correlation functions on a small cluster
to be more sensitive to symmetry than the  1-particle spectral
functions and the phase diagram, we use a symmetric arrangement of
the $O$ sites in the reference cluster. Additionally, the Coulomb
interaction $U_{pd}$ between holes on $Cu$ and $O$ sites is
implemented with periodic boundary conditions.

\begin{figure}
\begin{center}
  \includegraphics[width=0.5\columnwidth]{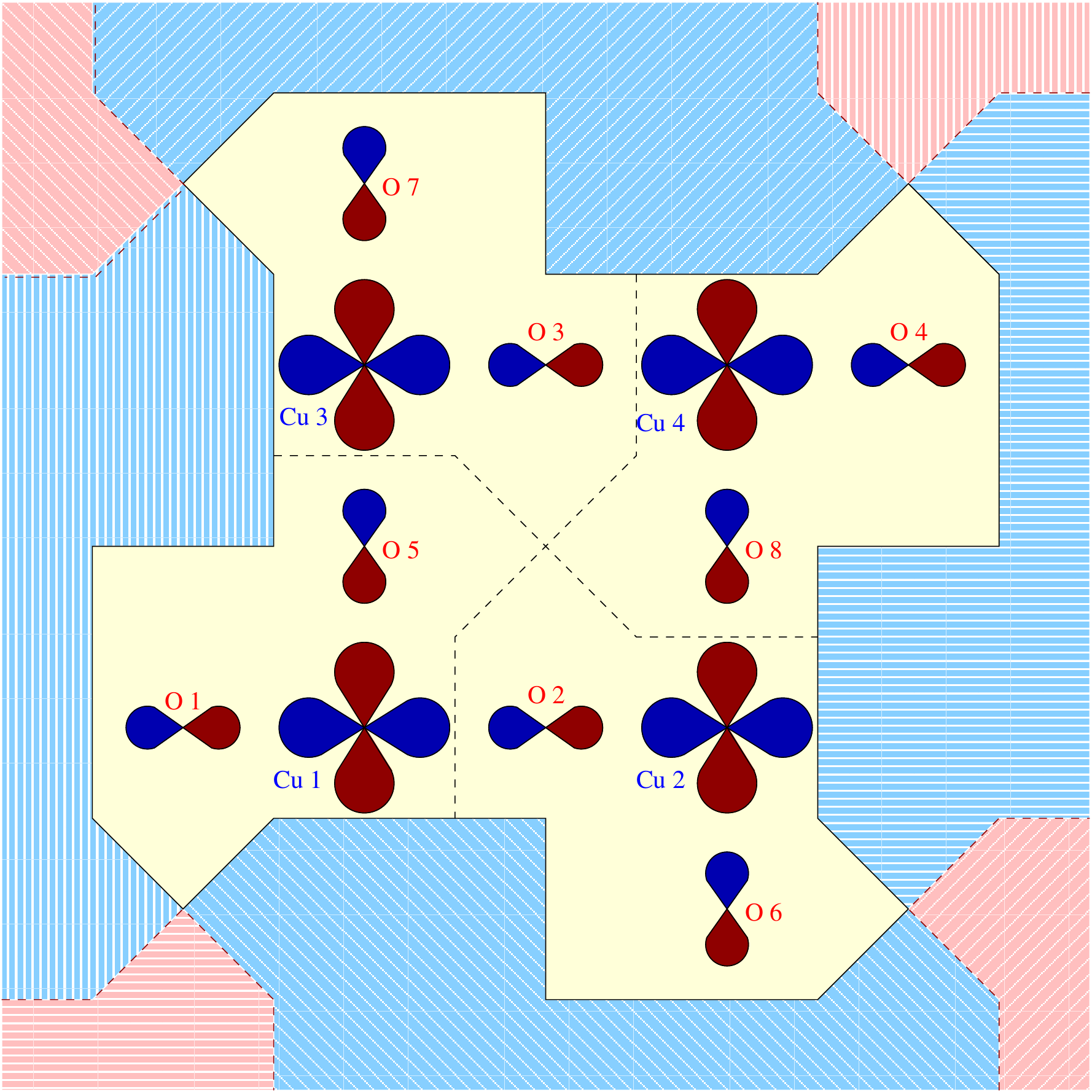}\\
  \caption{(2$\times$2) reference cluster used for 3BH-model. Each cell contains a $CuO_2$ unit. To reduce finite-size effects in 2-particle correlations, a more symmetric arrangement of $O$ sites is implemented, compared to previous publications (see Ref. \cite{ar.ai.09}).}
  \label{Figure 4}
\end{center}
\end{figure}

\begin{figure}
\begin{center}
  \includegraphics[width=0.75\columnwidth]{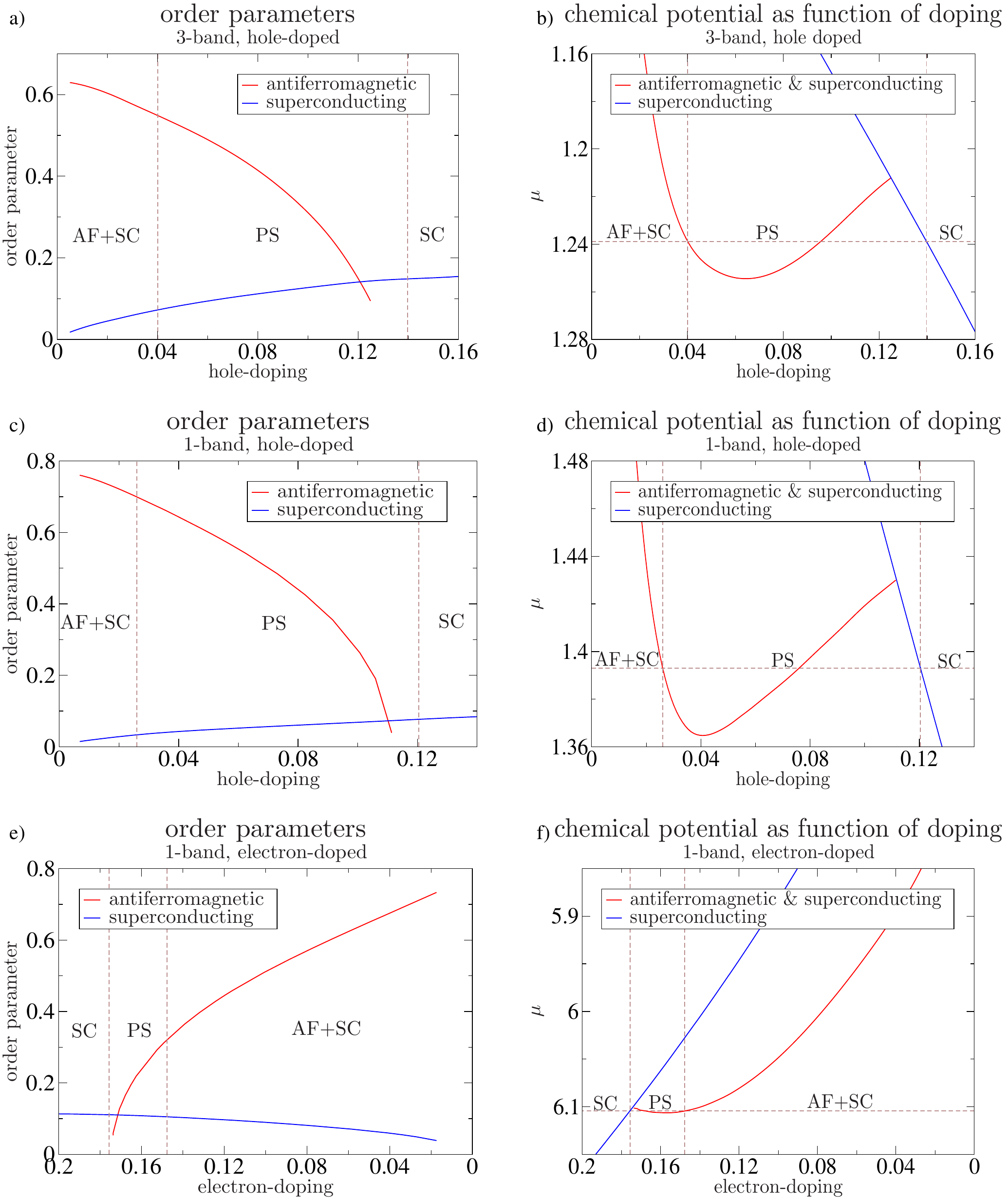}\\
  \caption{Ground-state phase diagram for the 3BH-model ( a) \& b) ), based on a (2$\times$2)
reference cluster with the parameters specified in the text, in comparison with the
1BH-model ( c) - f) ), based on a (4$\times$2)
reference cluster. In d), e) \& f), the chemical potential $\mu$ is plotted as a function of doping. The
corresponding order parameters for the models are given in (~a), c) \& e)~). Between the mixed phase with both antiferromagnetic and superconducting order (AF+SC) and the pure superconducting phase (SC) at higher-dopings, is a region of phase separation (PS), described in text. For 1BH-model, both hole- and electron-doping is presented.
There is no qualitative difference between 1BH and 3BH phase diagrams.}
  \label{Figure 5}
\end{center}
\end{figure}

\begin{figure}
\begin{center}
  \includegraphics[width=1\columnwidth]{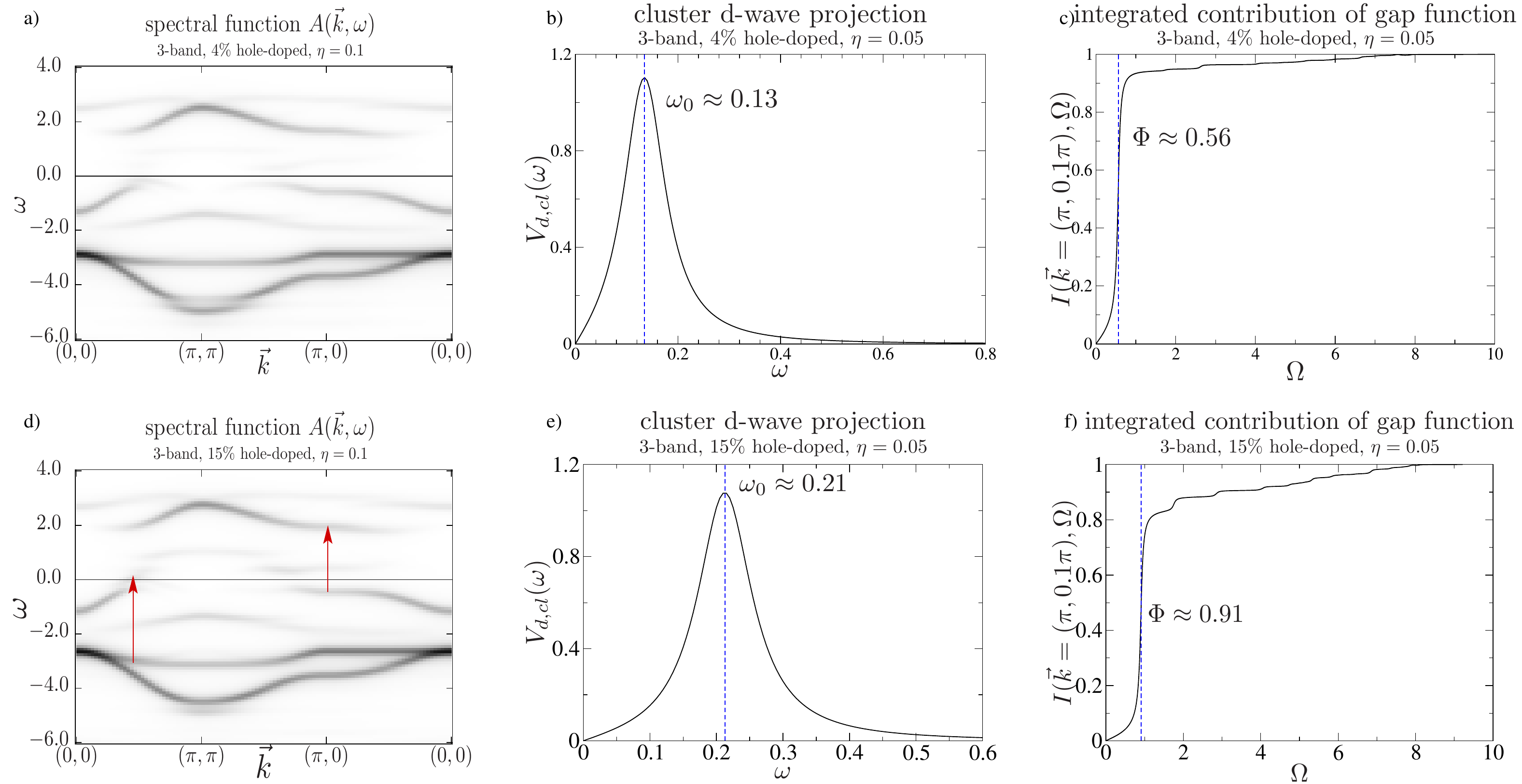}\\
  \caption{\textbf{3BH-model:} single-particle spectral function $A(\vec{k},\omega)$ (~a)~\&~d)~), d-wave projection $V_d(\omega)$ of the spin susceptibility (~b)~\&~e), see eq.~(10)) and integrated contribution of gap function $I(\vec{k}=(\pi,0.1\pi),\Omega)$ (~c)~\&~f)~) for a (2$\times$2) reference cluster (as in Fig. 4). For a)~-~c), the hole-doping is $4\%$ and for d)~-~f), the hole-doping is $15\%$. For both dopings, the system is in a pure superconducting state.}
  \label{Figure 6}
\end{center}
\end{figure}

On the other hand, the results for the 3BH-model phase diagram in
Figs.~\ref{Figure 5} and the 1-particle excitations in
Figs.~\ref{Figure 6}~a)~\&~d) are in close agreement with our
earlier results published in Ref. \cite{ar.ai.09}, where we used a
(not-symmetrized) (2$\times$2) cluster. The ``upshot'' of these
results is, that the 3BH- and the 1BH-model phase diagram results
with the latter presented in Figs.~\ref{Figure 5}~a)~\&~c) are very
similar, reproducing in both cases the overall ground-state phase
diagram of the high-$T_c$ superconductors (see also Fig.~3 of Ref.
\cite{ar.ai.09}). In particular, they include salient features, such
as the enhanced robustness of the AF state as a function of electron
doping and the tendency towards phase separation (``PS'' regime)
into a mixed AF-SC phase at low doping and a pure SC phase at high
(both hole and electron) doping. In the low-doping regimes, we find
at a homogeneous symmetry-broken state in which both, the AF and the
d-wave SC order parameters $m$ and $\Delta$ are nonzero. This
corresponds to a phase ``AF+SC'', where AF and d-wave SC order
microscopically coexist. A homogeneous state with pure d-wave SC
($m=0$ and $\Delta>0$) is obtained in the larger doping regimes. In
the ``in between doping'' regions macroscopic phase separation
occurs, where these two latter phases are thermodynamically
unstable. In previous work, we have shown for the 1BH-model, that
larger reference clusters still yield a qualitatively similar phase
diagram, with one noteable exception: the larger clusters (up to 10
sites in the 1BH-model) results suggest, that phase separation
remains persistent in  the h-doped case and disappears in the
$e$-doped situation \cite{25a}. Based on the overall very similar
phase diagrams for the (2$\times$2) 3BH-model
 and for the (4$\times$2) 1BH-model (Figs.~\ref{Figure 5}~b)~\&~d)),
we expect a similar disappearance of the ``PS''-region for $e$-doping
also in the 3BH-model.

\begin{figure}
\begin{center}
  \includegraphics[width=1\columnwidth]{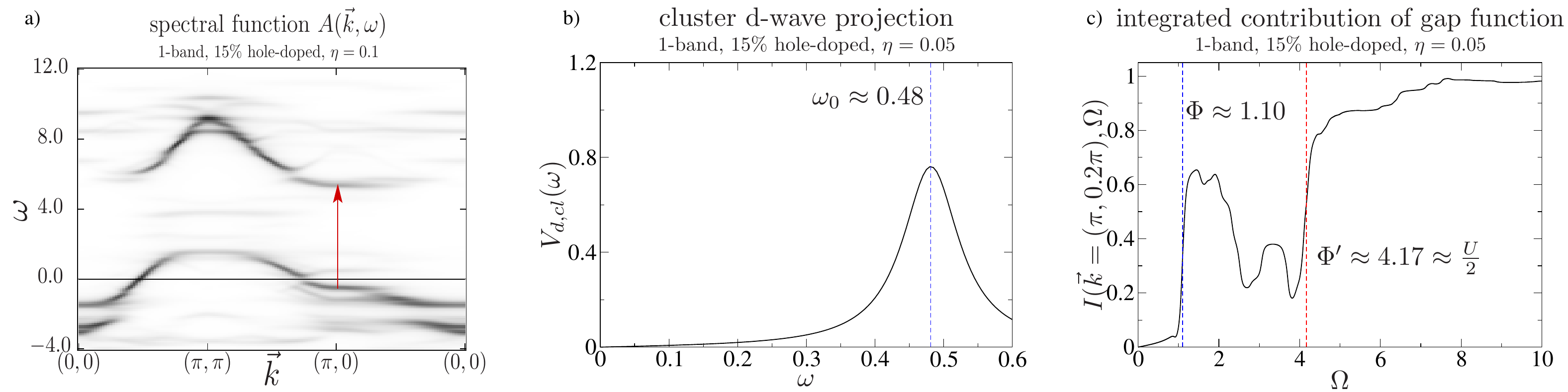}\\
  \caption{\textbf{1BH-model:} single-particle spectral function $A(\vec{k},\omega)$ (~a)~), d-wave projection $V_d(\omega)$ of the spin susceptibility (~b), see eq.~(10) ) and integrated contribution of gap function $I(\vec{k}=(\pi,0.2\pi),\Omega)$ (~c)~) with a (3$\times$3) reference cluster. Hole-doping is $15\%$ and the system is in a pure superconducting state (see Fig.~\ref{Figure 6}).}
  \label{Figure 7}
\end{center}
\end{figure}

Why are the phase diagrams in Fig.~\ref{Figure 5} so similar despite
the fact that, for example, the p-d charge fluctuations give rise to
an at higher energies $(\omega \sim \Delta_{pd})$ even qualitatively
different behavior of the dynamic pairing interaction for the 3BH
results when compared to the 1BH-data as shown in Figs.~\ref{Figure
6}~\&~\ref{Figure 7} for $I(\vec{k}, \Omega)$? The reason is that on
the scale of the (maximal) d-wave gap energies, $\Delta_{pd}$ is a
doping-independent ``high-energy'' scale ($\sim$ ten times larger
than the SC gap) the role of which is taken over in the 1BH-model by
another doping independent ``high-energy'' scale, namely an
effective Hubbard $U$ (see discussions, below). As just stressed,
however, the dynamics of the pairing mechanism, may be quite
different, which is what we find indeed below. We expect this to be
of importance for the "non-universality", i.e. material-dependence
of the cuprate SC.

Also the single-particle excitations (as displayed in Figs.~6 up to
7 as well as in our earlier Ref. \cite{ar.ai.09}) are found
qualitatively to be similar, concerning the ``low-energy'' physics
and, in particular, the $e$-$h$ asymmetry. This asymmetry explains
the corresponding $e$-$h$ asymmetry found in the ground-state phase
diagram for the robustness of the AF phase (Figs. \ref{Figure
5}~c)~\&~e): doped holes first enter around the nodal point
($\frac{\pi}{2},\frac{\pi}{2}$), where the SC gap vanishes,
introducing a ``gap-less'' screening which very effectively destroys
long-range AF order \cite{ref:16}. In contrast, introducing
electrons around the anti-nodal point ($\pi,0$) fixes $\mu$ (due to
the large density of states) within the SC gap, in a regime where
this gap is maximal. Here, an incomplete screening cannot disrupt AF
order up to significantly large $e$-dopings where, finally, $\mu$
enters also here the ``gap-less'' ($\frac{\pi}{2},\frac{\pi}{2}$)
region. So again, it is the corresponding low-energy physics
embedded in the qualitatively rather similar single-particle spectra
plotted in Figs.~6~-~7, which determines salient (here AF) features
of the phase diagram.

\emph{So, let us finally come back to the question of the dynamics
in the pairing interaction and the fractional contribution
$I(\vec{k}, \Omega)$ to the gap function in the 3BH- as well as 1BH-
models.}

Our results for $I(\vec{k}, \Omega)$ for the 3BH-model, i.e. for the
fraction of the zero-frequency gap function which arises from
frequencies below $\Omega$, are plotted in Figs.~\ref{Figure
6}~c)~\&~f). Here two doping cases, $4\%$ hole doping and $15\%$
hole doping are shown, together with the corresponding
single-particle spectral function $A(\vec{k}, \omega)$ and the data
for the d-wave projection of the cluster spin susceptibility (eq.
(10)).

$I(\vec{k}, \Omega)$ is plotted for the 3BH-model for  a doping
$x=4\%$ in Fig.~\ref{Figure 6}~c) and for a higher doping $x=15\%$
in Fig.~\ref{Figure 6}~f). $\vec{k}=\pi(0.1,0)$, i. e. a
$\vec{k}$-point close to the anti-nodal point ($\pi$,0) where the
d-wave SC gap is maximal. For both dopings, we observe a first steep
rise at a typical energy $\Phi$ of about $\omega_0+\Delta_0$, where
$\Delta_0$ denotes the quasi-particle (SC) gap and $\omega_0$ a
characteristic spin-fluctuation frequency. Close to half-filling,
$\omega_0$ is roughly estimated by the strong-coupling result for
the $Cu$-$Cu$ exchange energy $\omega_0 \cong 2J_{Cu}$, with

\begin{equation}
J_{Cu}=
\dfrac{4t_{pd}^4}{(\Delta_{pd}+U_{pd})^2}(\dfrac{1}{\Delta_{pd}+U_{pp}/2}+\dfrac{1}{U_{dd}})\cong 0.1t_{pd}.
\end{equation}

As shown by the d-wave projected result $V_{d,cl}(\omega)$ in
Fig.~\ref{Figure 6}~b)~\&~e), for the imaginary part of the cluster
spin susceptibility, with the definition

\begin{equation}
V_{d,cl}(\omega) = \frac{1}{N^2} \sum \limits_{\vec{k},\vec{k}'} \left( \cos(k_x) -
\cos(k_y) \right) \left( \cos(k_x') - \cos(k_y') \right) \text{Im}
\chi_{cl}(\vec{k}-\vec{k}',\omega),
\end{equation}

the first dominant peak is found at low doping, i.e. $4\%$ at about
this energy (here $\omega_0 \cong 0.13$ as always in units of the
hopping $t_{pd}$) and, at $15\%$ doping at a higher energy
$\omega_0\cong0.2$.

$I(\vec{k}, \Omega)$ displays an amazingly prominent first increase
at an energy $\Phi$ of about $\Delta_0+\omega_0$, where $\Delta_0$
is the quasiparticle gap and $\omega_0$ the first dominant peak in
the d-wave projection of the dynamic spin susceptibility
$\text{Im}\chi(\vec{k},\omega)$. The quasiparticle gap is found to
be at $\Delta_0\cong0.52$ (obtained from the spectral function) for
$4\%$ doping and at a similar value ($\Delta_0\cong0.41$) for $15\%$
doping, thus $\Delta_0+\omega_0\cong 0.65$. The first sharp rise in
$I(\vec{k},\Omega)$ at the low-doping situation ($\sim 4\%$) is at
about $\Phi\cong0.56$. We observe that in this low-doping case for
the 3BH-model more than $90\%$ of the saturation value of
$I(\vec{k},\Omega)=1$ is due to the dynamic contribution of the spin
fluctuations.

\begin{figure}
\begin{center}
  \includegraphics[width=1\columnwidth]{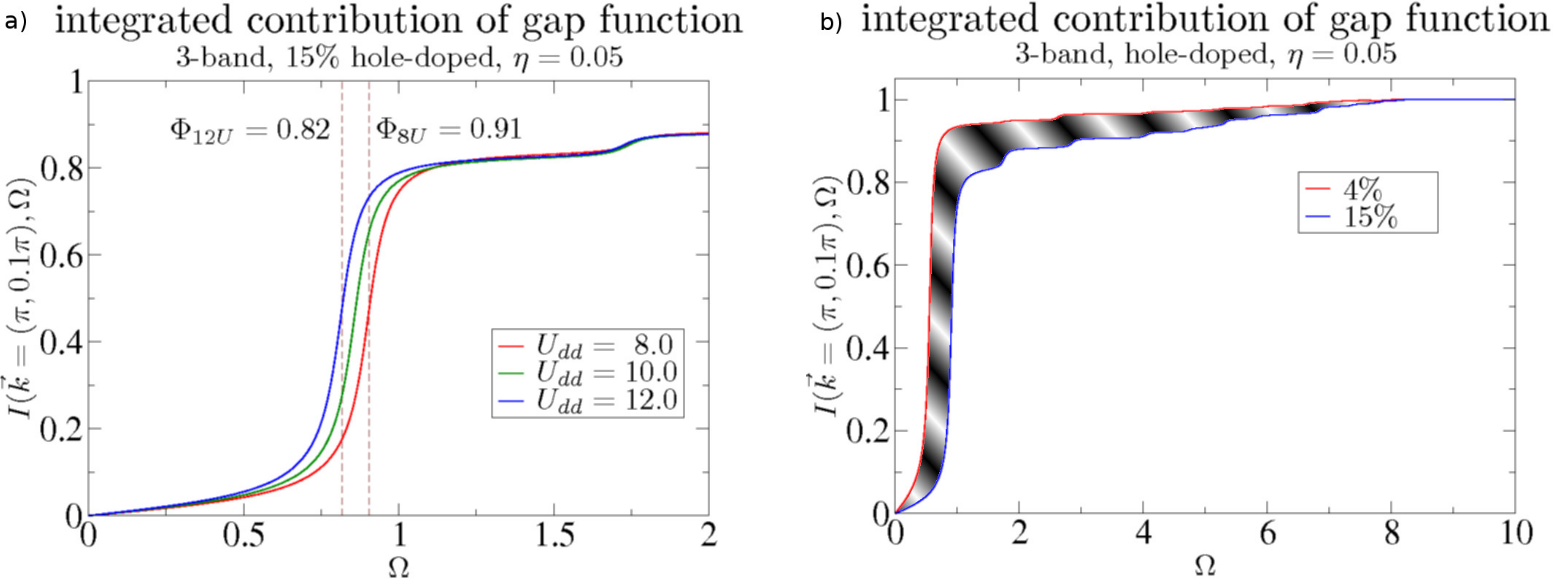}\\
  \caption{Integrated contribution of gap function $I(\vec{k}=(\pi,0.1\pi),\Omega)$ for the 3BH model.
  In a), the first dominant increase in $I(\vec{k},\Omega)$ is plotted for different values of $U_{dd}/t$. In b), the doping dependence of $I(\vec{k},\Omega)$ is shown, with the modest increase of p-d charge fluctuations as function of doping.}
  \label{Figure 8}
\end{center}
\end{figure}

For $15\%$ doping $\omega_0$ is increasing \cite{ref:13}, and so is
$\omega_0+\Delta_0$. Fig.~\ref{Figure 8}~a) further confirms the
leading role of the spin-fluctuations. Here the first dominant
increase in $I(\vec{k},\Omega)$ is plotted as a function of $U/t$ at
15$\%$ h-doping. One sees that as the value of $U/t$ increases, this
dominant increase in $I(\vec{k},\Omega)$ is shifted to lower
energies, scaling essentially with the exchange energy $J(\Delta_0$
is found to be essentially constant as function of $U/t$) Here, at
$15\%$ doping, the spin-fluctuation contribution to the saturation
value of $I(\vec{k},\Omega)=1$ slightly decreases compared to the
$4\%$-doping situation. From there on a mild rise takes place,
starting at about $\omega\cong1$ for both dopings. In both doping
cases this corresponds to virtual e-h transitions involving the
charge transfer gap $\sim\Delta_{pd}/2$ (see Fig. \ref{Figure 7}a).

Fig.~\ref{Figure 8}~b) illustrates the influence of the doping
dependence of p-d charge fluctuations on $I(\vec{k},\Omega)$. Here,
at first glance, the influence of these fluctuations on the
integrated gap function $I(\vec{k},\Omega)$ appears to be very mild.

However, this picture changes significantly, when we confront the
3BH-model results for $I(\vec{k},\Omega)$ in Figs. \ref{Figure
6}~d)~-~f) with the corresponding ones for the 1BH-model and, thus,
the doped Mott-insulator case (Figs. \ref{Figure 7}~a)~-~c): again,
we find for the latter model a first dominant rise in the integrated
gap function. (We plot here only the results for the $15\%$-doping
case with the low-doping ($4\%$) results being qualitatively
similar.) This first rise - when analyzing its ($U/t$)-dependence -
can again be attributed to dynamic spin excitations. However, a
second prominent increase in this 1BH-model case occurs after a
rather pronounced drop at energies of order $(U/t)/2$. This has
already been observed in this 1BH-model case in Ref. \cite{ref:26}
and was termed a ``non-retarded'' contribution occuring at an energy
scale set by the Mott gap and being related to excited states (see
the red arrow in the corresponding $A(\vec{k}, \omega)$ spectrum in
Fig.~\ref{Figure 7}~a) involving the upper Hubbard band. For the
$t$-$J$ model, Ref \cite{ref:26} finds this energy scale being
pushed to ``infinity'', with the corresponding exchange interaction
being instantaneous. In contrast, the retarded (spin-fluctuation)
interaction , i.e. the first steep rise in $I(\vec{k},\Omega)$,
occurs at an energy scale which is small ($\omega_0\cong0.1t$)
compared to the bare bandwidth $8t$ (and the bare $U$). Thus, the
\emph{1BH-model as well as the $t$-$J$} model may be interpreted to
\emph{contain both retarded and non-retarded contributions}.

This is \emph{qualitatively different in the 3BH-model}: Here, in
Figs.~\ref{Figure 6}~c)~and~f) and Fig.~\ref{Figure 8}~b) we see a
rather continuous ``filling in'' of integrated weight in the
$\Omega$ regions in $I(\vec{k},\Omega)$, where in the 1BH-model
(Fig.~\ref{Figure 7}~c) ) this weight is clearly missing. This
``filling in'' is due to electron-hole excitations of $\mathcal
O(2t_{pd})$ and, thus, due to p-d charge fluctuations (see the
corresponding ``red arrow'' for the $e$-$h$ transitions in
$A(\vec{k},\omega)$).

Thus, in the sense of the interpretations used for the 1BH-model
above (and in Ref \cite{ref:26}) we may conclude that the
\emph{3BH-model is indeed different and contains only retarded
contributions in its d-wave pairing interaction}.

In summary, then, the question posed in sect. III, i.e. whether
there is a ``pairing glue'' in the 3BH-model is a question of
whether the dominant contribution to the pairing function
$\phi_1(\vec{k},\omega=0)$ comes from the integral of
$\phi_2(k,\omega)/\omega$  and, more specifically, a ``low-energy''
(compared to 8t) region in this integral. Earlier results have shown
that both the 1BH- and t-J models exhibit spin-fluctuation ``pairing
glue'' \cite{ref:26,ref:31}. For both the 3BH-model as well as for
the 1BH-model results, with the latter displayed for a similar
higher-doping case ($15\%$) in Fig.~\ref{Figure 7}, one can include
the full range of virtual $e$-$h$ transitions between the lower and
upper Hubbard bands. As a consequence the dispersion (Cauchy)
relation for $\phi(\vec{k},\omega=0)$ does, strictly speaking, not
contain a non-retarded contribution $\phi_{\text{NR}}$ and the
asymptotic value of $I(\Omega)$ approaches unity. In the 1BH-model,
we observe a first steep rise, which can again be identified with a
typical spin-fluctuation energy. Then, at higher energies after a
drop a second steep rise occurs at energies corresponding to
(virtual)e-h interband transition of order $\sim\frac{U}{2}$. As
argued in Ref. \cite{ref:26}, it is this part which in the t-J model
for $U\rightarrow\infty$ (i. e. when the upper Hubbard band is
projected out) gives rise to an instantaneous contribution
$\phi_{\text{NR}}$. Thus, in this sense, both 1-band models also
display a non-retarded interaction.

Our new result here is that \emph{in the 3BH-model only low-energy
(spin-fluctuation) retarded contributions to the pairing interaction
dominate, whereas in the 1BH-model also a significant high-energy
contribution occurs}. While these differences between a Mott and a
charge-transfer insulator appear to be renormalized away in the very
similar, i.e. ''universal'' ground-state phase diagram of these two
models, one may use Eliashberg-type of arguments to form an
expectation that the finite-T phase is rather different in the two
models, i.e. strongly material-dependent (see also Ref.
\cite{ref:25}).




\section{Spin susceptibility}
\newcommand{\vvv}[1]{\mbox{\boldmath{$#1$}}}
\newcommand{\beq}{\begin{equation}}
\newcommand{\eeq}{\end{equation}}
\newcommand{\beqn}{\begin{eqnarray}}
\newcommand{\eeqn}{\end{eqnarray}}
\newcommand{\qaf}{Q_{AF}}
\newcommand{\eqqref}[1]{Eq.~(\ref{#1})}

As is clear from the previous sections, two-particle (2-p)
excitations, especially magnetic ones, are fundamental for
understanding the issue of the ``pairing glue'' in high-T$_c$
cuprate superconductors. In addition, for hole-doped cuprates in the
superconducting (SC) state, the celebrated resonant magnetic mode
emerges with its peak intensity being highest around the wave vector
$\qaf=(\pi,\pi)$ \cite{pa.si.04}. Away from $\qaf$, the mode has
both a downward and upward ``hour-glass''-like dispersion. 
For electron-doped materials~\cite{wi.da.06}  the magnetic
excitation spectrum has a different structure: it is confined to a
small momentum region around $(\pi,\pi)$ and it is essentially
dispersionless \cite{is.er.07}.

In order to provide an appropriate description of the magnetic
susceptibility and, in particular, of the magnetic resonance mode it
is important to adopt a theory which is working in the
experimentally relevant strong-correlation regime. As discussed
above, VCA provides a suitable tool to study strongly-correlated
systems for large lattices.  In Ref.~\cite{br.ar.10} we have
developed a novel theory for two-particle excitations based on the
VCA. This approach is, in contrast to previous weak-coupling and/or
semiphenomenological treatments, parameter free and applies to the
relevant strong-coupling regime of the one- and three-band Hubbard
models.
Within this method, two-particle susceptibilities are evaluated
 by first computing the {\it exact} two-particle irreducible vertex
$\vvv \Gamma$ within the reference system (cluster), and then
inserting  $\vvv \Gamma$ into the Bethe-Salpeter (BS) equation for
the exact (lattice) two-particle susceptibility~\cite{br.ar.10}.
Notice that the BS equation is in principle exact and not restricted
to weak-coupling (in contrast to RPA). In principle, $\vvv \Gamma$
depends on three frequencies and four cluster indices. Therefore,
solving the BS equation with the corresponding full frequency and
momentum dependence of $\vvv \Gamma$ is a numerically quite
expensive task. For this reason, we  replace the full $\Gamma$ with
its average over external frequencies, and set the site indices
pairwise equal, which corresponds to averaging over the external
cluster momenta. After this simplification, $\vvv \Gamma(i \omega)$
can be computed as \cite{br.ar.10,ref:45}
\beq \vvv\Gamma(i \omega)
= \left[(\vvv
  \chi^{0'}(i \omega))^{-1} - (\vvv
  \chi'(i \omega))^{-1} \right] \;,
\eeq
where $\vvv \chi'$ is the exact  and $\vvv \chi^{0'}$ is the
``bubble''  susceptibility of the reference system, and $\omega$ is
a Matsubara frequency.
 The  term $\vvv \chi'$ in eq.~(11) is evaluated directly within a Lanczos
procedure and the first one is computed
 using the exact cluster Green's functions.
The effect of the above approximations is then partially compensated by
 multiplying
$\vvv \Gamma(\vvv Q,i \omega)$ by a constant term $\alpha$ which is
determined so as to fulfill  the sum rule for the transverse spin
susceptibility. Details can be found in  Ref.~\cite{br.ar.10}.
Finally, the lattice spin (or, corresponding charge) susceptibility,
obtained by solving the BS equation, is given by \beq
 \vvv \chi(\vvv Q, i \omega))^{-1} =
 \vvv \chi^0(\vvv Q, i \omega))^{-1} -
\alpha \vvv\Gamma(i \omega) \;, \eeq where  $\vvv \chi^0(\vvv Q, i
\omega))^{-1}$ is the lattice ``bubble'' susceptibility evaluated
with the fully dressed lattice Green's functions. Since we are
studying the superconducting phase, the bubble susceptibility has a
contribution from both normal and anomalous Green's
functions.~\cite{br.ar.10} Due to the partial breaking of
translation symmetry introduced by the cluster tiling, both the
$\chi_0$ and $\chi$  depend on the wave vector $\vvv Q$ of the
correspondingly reduced Brillouin zone, and are matrices in Nambu
space as well as in the cluster sites.

\begin{figure}[h]
\begin{center}
 \includegraphics[width=0.8\textwidth]{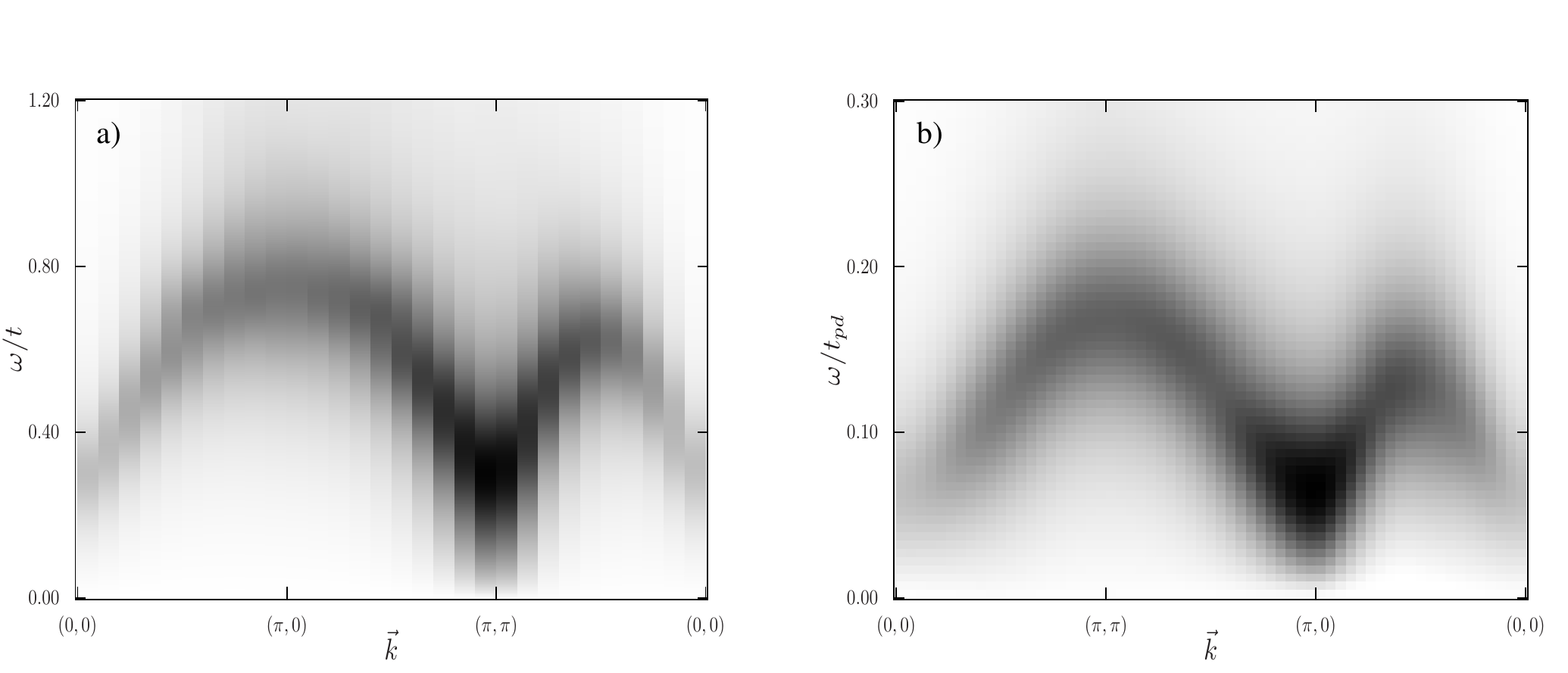}
\caption{
 Imaginary part of the spin susceptibility
    $\chi^{\pm}(\vec{k},\omega)$ for the single-band ( a) ) and for the
    three-band ( b) ) Hubbard model at doping $x=4\%$. Results are
    obtained with a reference system of (2$\times$2) unit cells.
\label{spin-compare}
}
\end{center}
\end{figure}

In Ref.~\cite{ar.ai.09}, as well as in the present paper, the
qualitative similarity of the phase diagram and  single-particle
excitations between the three-band  and the  single-band  Hubbard
models was demonstrated. In particular, it was explicitly confirmed
that the asymmetry between  electron- and hole-doped cuprates,
despite being of fundamentally different nature, shows very similar
signatures in the single-particle spectrum of both models, provided
a
 next-nearest-neighbor hopping
$t'$, is included in the latter.
 The question remains whether this similarity can be extended to two-particle
excitations.
To address this question, we compare in Fig.~\ref{spin-compare}
 the
imaginary parts of the spin susceptibility evaluated in the deeply
underdoped regime ($x=4\%$), for both the single- and the three-band
models.

While there are some minor differences, both spectra display a
dispersion which is a remnant of the spin-wave band in the
antiferromagnetic half-filled phase. This fact signals the presence
of strong antiferromagnetic fluctuations at this doping
concentration also in a doped charge-transfer insulator.

\begin{figure}
\begin{center}
 \includegraphics[width=0.8\textwidth]{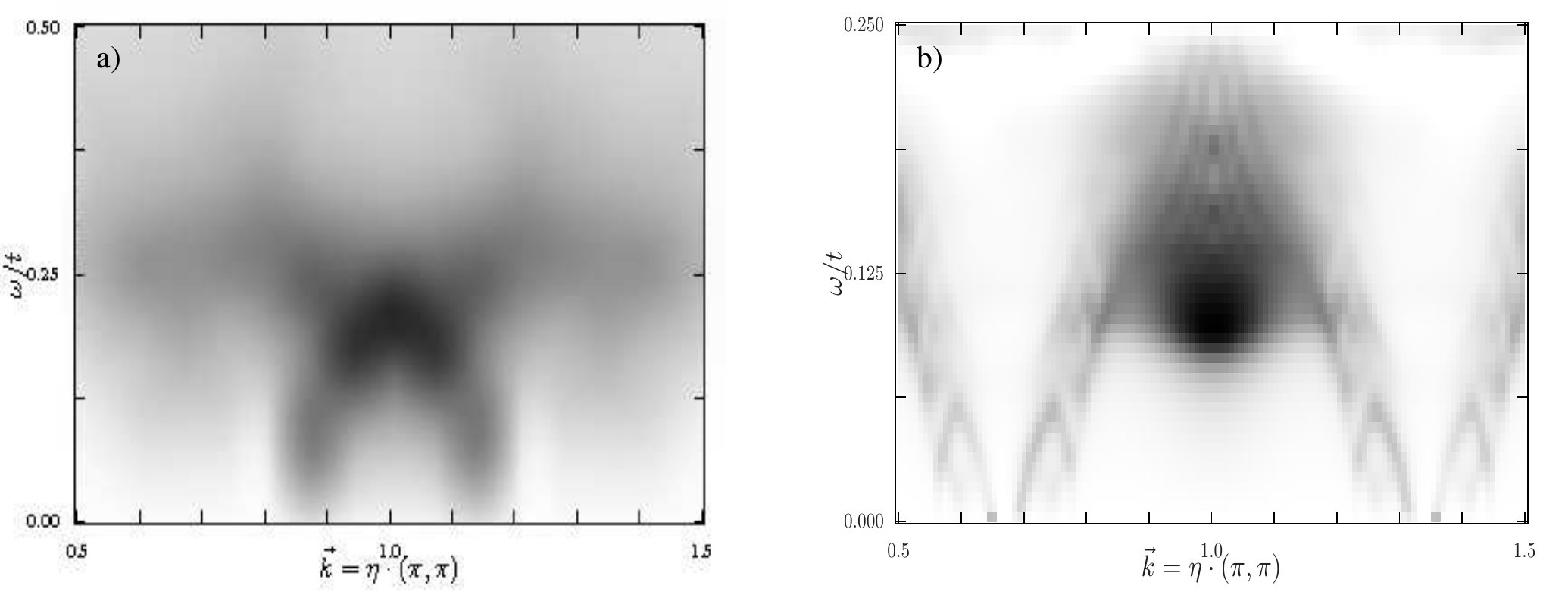}
\caption{
 Imaginary part of the spin susceptibility
 for the single-band Hubbard model ((3$\times$3) reference clusters, $U=8t$)
in the hole-doped ($x=0.15$, a) , taken from Ref.~\cite{br.ar.10} ) and in the electron-doped ($x=0.14$, b) )
cases.
\label{el-ho}}
\end{center}
\end{figure}

Last not least, the \emph{VCA-extension to 2-particle
susceptibilities can account for} - without any adjustable
parameters -  \emph{the magnetic resonance} in both hole- and
electron-doping regimes. Close to optimal doping, the magnetic
spectrum for the h-doped single-band Hubbard model
(Fig.~\ref{el-ho}~a)~) shows the famous magnetic resonance peak at
energies of about $0.1t - 0.2t$ and around wave-vector $(\pi,\pi)$.
From this resonance, a downward dispersion extends down to the onset
of the particle-hole continuum. The striking feature leading to  the
celebrated ``hourglass'' structure is, however, the additional
``upward'' dispersion, which can be seen in the spectrum of the
single-band Hubbard model displayed in Fig.~\ref{el-ho}~a). Because
of computational difficulties, we were not able to obtain this
feature in the three-band model calculation. This is due to the fact
that the smaller (2$\times$2) $CuO_2$-units cluster adopted for the
three-band model still has significant antiferromagnetic
correlations which prevent the formation of the hourglass structure
(a similar observation holds also for the 1BH-model, where we had to
go to (3$\times$3) clusters (Fig.~\ref{el-ho}) to be able to observe
the resonance).

In the electron-doped case, the magnetic excitation spectrum
displays a different structure. In Fig.~\ref{el-ho}~b), we present
the first strong-coupling and parameter-free evaluation of the
magnetic spectrum in the electron-doped case, within our VCA
approach for two-particle excitations. As one can see, in this case
the weight of the magnetic spectrum is concentrated essentially in
the vicinity of $(\pi,\pi)$, in accordance with the experimental
situation. The downward dispersion observed in the calculation only
carries very little weight, when compared with the strong peak
around $(\pi,\pi)$. Therefore, this might be the reason why this
dispersion is
 not seen in inelastic neutron scattering
 experiments~\cite{wi.da.06}. Finally,
 in contrast to the hole-doped case, the upward dispersion
is not observed for electron doping. This is also in accordance with
the experimental observations \cite{is.er.07}.

We have also extended, in a similar spirit, the dynamical cluster
approximation (DCA) to 2-p susceptibility calculations at
finite-temperatures \cite{ref:45} for the infinite-lattice case. In
Ref. \cite{ref:45}, this scheme has been used to obtain - in  the
case of the 1BH-model - information on the finite-T phase diagram
(via the instability towards a symmetry-broken phase of the
corresponding susceptibility).

\section{Conclusion}
One key issue in the research on high-$T_c$ cuprate
superconductivity is to derive a detailed understanding for the
specific role of the $Cu$-$d_{x^2-y^2}$ and $O$-$p_{x,y}$ orbital
degrees of freedom. Very early on a corresponding three-band Hubbard
(3BH) model has been suggested but in most studies an effective
one-band Hubbard (1BH) or $t$-$J$ model has been used. The one-band
models have been shown, using a variety of techniques, to reproduce
salient features of the cuprates such as the competing
antiferromagnetic (AF) and superconducting (SC) phases including the
electron-versus-hole doping asymmetries. Here, the physics is that
of doping into a Mott-Hubbard insulator, whereas the actual
high-$T_c$ cuprates are doped charge-transfer insulators. Using a
cluster embedding scheme, i.e. the variational cluster approach
(VCA) which allows for a controlled route to go to the large system
size limit (and, thereby, low-energy limit), we compare in detail
the competing phases in the ground-state, the single-particle and
two-particle excitations and, in particular, the dynamic pairing
interaction and the issue of a ``pairing glue'' of the 3BH and 1BH
model. This study demonstrates that there are pronounced differences
on a ``higher-energy'' scale (compared to the SC gap energy) for the
two models, such as the charge-transfer energy $\Delta_{pd}$ in the
3BH-model. While these ``high-energy'' features,where the
material-dependent physics enters, play no decisive role for the
overall structure of the phase diagram, which shows excellent
agreement between 3BH- and 1BH-models, there appear significant
differences in the dynamics of the pairing interaction: in the
3BH-model the interaction is dominated by a retarded pairing due to
low-energy spin fluctuations, whereas in the 1BH-model in addition a
part comes from excited states involving the upper Hubbard band,
which may be associated with a ``non-retarded'' contribution. Thus,
one may term the ground-state phase diagram ''universal'', whereas
the different dynamics of the pairing interaction of the two models
should be reflected in a pronounced material dependence of the
finite-T phase diagram.

First steps in this direction have already been undertaken in Ref.
\cite{ref:25} and are also presently followed up in our group
\cite{ref:46}.

Finally a comment about the RVB-type of physics in the two models:
$\Delta_{pd}$ also enters in the exchange coupling $J$ between spins
on the $Cu$ sites. Thus, the role of $U$ in the 1-band model,
setting the characteristic energy of the RVB-coupling $J$, is taken
over (see Eq. (9)) in the 3-band model by the energy of the $p-d$
charge fluctuations. In that sense, one may argue, that the final
increase in the relative pairing strength $I(\vec{k}, \omega)$ to
the value of $I(\vec{k}, \omega)=1$ again reflects RVB physics.
However, our results demonstrate that there is quite a difference
with respect to the relative weights of the spin-fluctuation and
higher-energy contributions between the two models.

\section*{Acknowledgments}
We acknowledge detailed discussions with D.J. Scalapino, Th. Maier
and M. Potthoff and financial support form the DFG HA1537/21-3, from
the Bavarian supercomputer network KONWIHR and from the FWF
P18551-N16.

\end{document}